\def\la{\;
\raise0.3ex\hbox{$<$\kern-0.75em\raise-1.1ex\hbox{$\sim$}}\; }
\def\ga{\;
\raise0.3ex\hbox{$>$\kern-0.75em\raise-1.1ex\hbox{$\sim$}}\; }
\documentstyle[psfig,conf_iap,11pt]{article}
\begin{document}
{\small 
\heading{How accurately can the deuterium abundance be determined ?
} 
\par\medskip\noindent
\author{ Sergei A. Levshakov$^1$, Wilhelm H. Kegel$^2$ and 
Fumio Takahara$^3$}
\address{Department of Theoretical Astrophysics, A. F. Ioffe Physico-Technical Institute,
194021 St. Petersburg, Russia }
\address{Institut f\"ur Theoretische Physik der Universit\"at Frankfurt am Main, Postfach 
11 19 32, 60054 Frankfurt/Main, Germany}
\address{Department of Physics, Tokyo Metropolitan University,
Hachioji, Tokyo 192-03, Japan}

Modern observations of the H+D absorption at high redshift play an important
role in determining the primordial hydrogen isotopic ratio 
D/H $\equiv N_{\rm DI} / N_{\rm HI}$ (the ratio of the DI to HI column densities).
Measurements of $N_{\rm DI}$ and $N_{\rm HI}$ from QSO spectra can provide a sensitive
test of the predictions of big bang nucleosynthesis [BBN] if their precision is
comparable with the theoretical uncertainties ($\simeq 15$ \%).

It has been shown, however, that the determination of $N$ depends on the assumptions made
with respect to the line broadening mechanism and 
may be ambiguous especially for the
case of optically thick lines [1--8]. The commonly used procedure is to fit Voigt profiles,
accounting for radiative damping and Doppler broadening in the {\it microturbulent} limit
(completely uncorrelated bulk motions). But the interpretation
of observed profiles may be substantially changed if a {\it mesoturbulent} model is
used, i.e. if the influence of a {\it finite}  correlation length in the stochastic
velocity field on the line forming process is accounted for. This appears to be 
appropriate in particular with respect to the H+D absorption observed in QSO spectra.
The hydrogen
absorbers with $N_{\rm HI} \sim 10^{17} - 10^{18}$ cm$^{-2}$ (which are especially useful
for deuterium observations) are usually associated with extended gas halos of
foreground galaxies. Direct observations \cite{VO} of distant galaxies ($z > 2$)
reveal complex morphologies and kinematics of the extended HI Ly$\alpha$ absorption gas 
with projected sizes up to $\sim 50$ kpc. The inferred high values of the rms velocity
dispersion ($\sigma_t \sim 20 - 50$ km s$^{-1}$) and a few kpc sizes of the absorbing
regions make the mesoturbulent approach more adequate in this case.

Within the framework of the mesoturbulent model, one has to determine both
the physical parameters of the gas and the velocity field structure $v(s)$
along the line of sight. This problem may be successfully solved by using
a Reverse Monte Carlo [RMC] technique \cite{L6}.

We applied the RMC procedure to a 
template H+D Ly$\alpha$ profile which reproduces the
Q1009+2956 spectrum with the DI Ly$\alpha$ line seen at $z_a = 2.504$ \cite{BT}.
Here we present only a part of our results, the full analysis is given in \cite{L6}.

To simulate real data, we added the experimental uncertainties to the
template intensities which were sampled in equidistant bins as shown in Fig.~1
by dots and corresponding error bars. A {\it one}-component mesoturbulent model
with a {\it homogeneous} density and temperature was adopted. Adequate profile fits
for three different sets of parameters are shown in panels
({\bf a1}, {\bf b1}, {\bf c1}) by solid curves, whereas the individual HI and DI
profiles are the dashed and dotted curves, respectively. The estimated parameters
(D/H, $N_{\rm HI}$, $T_{kin}$, the ratio of the rms turbulent velocity to the
thermal velocity $\sigma_t / v_{th}$, the ratio of the cloud thickness to the
correlation length $L/l$) and $\chi^2_{min}$ values per degree of freedom are
also listed in these panels for each RMC solution.

The essential difference between the results of the {\it two}-component
microturbulent model adopted in \cite{BT} and ours lies in the estimation
of the hydrodynamical velocities in the $z_a = 2.504$ absorbing region.
The RMC procedure yields $\sigma_t \simeq 25$ km s$^{-1}$, whereas the
{\it two}-component model leads to $\sigma_t \simeq 2$ km s$^{-1}$ which
is evidently too low as compared with observations \cite{VO}.

In this particular absorption system, 
the study of the H+D Ly$\alpha$ profile yields D/H = $(3.75 \pm 0.85)\times10^{-5}$
$(2\sigma)$ which is slightly higher than the value $2.51^{+0.96}_{-0.69} \times 10^{-5}$
found in \cite{BT}. This difference is, however, very significant because it
leads to limits on D/H consistent with 
BBN predictions and observational constraints on both extra-galactic $^4$He mass
fraction Y$_p$ \cite{ITL} and $^7$Li abundance in the atmospheres of
population II (halo) stars \cite{BM}.
BBN restricts the common interval of the baryon to photon ratio $\eta$ for
which there is concordance between the above-mentioned abundances to
$(3.9 - 5.3)\times10^{-10}$ as shown by the shaded region in Fig.~2. It implies that 
$0.014 \la \Omega_b h^2_{100} \la 0.020$. 

We conclude that the discordance of D/H with the $^4$He and $^7$Li measurements
noted in \cite{BT} is a consequence of the use of the microturbulent model.
The generalized mesoturbulent model gives good agreement between the 
light element abundances and the BBN predictions.

\acknowledgements{This work was supported in part by the RFBR grant
No. 96-02-16905-a and by the Deutsche Forschungsgemeinschaft.}

\begin{figure}
\centerline{\vbox{
\psfig{figure=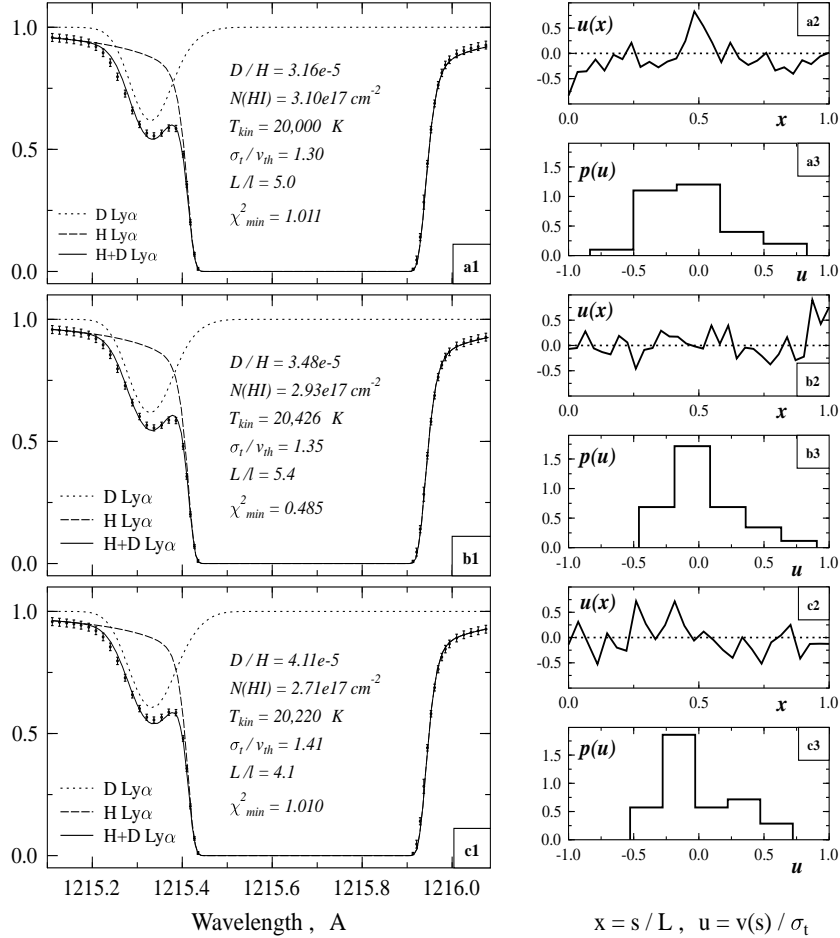,height=14.5cm,width=11.0cm}
}}
\caption[]{({\bf a1}, {\bf b1}, {\bf c1}) -- A template H+D Ly$\alpha$ profile
(dots with error bars) representing the normalized intensities and their
uncertainties in accord with \cite{BT}. The solid curves show the results of
the RMC minimization, whereas the dotted and dashed curves are the separate
profiles of DI and HI, respectively. Also shown are the best fitting parameters
and the $\chi^2$ values obtained for each case. ({\bf a2}, {\bf b2}, {\bf c2}) --
The corresponding individual realizations of the velocity distribution $u(x)$
in units of $\sigma_t$. ({\bf a3}, {\bf b3}, {\bf c3}) -- The histograms are
the projected velocity distributions $p(u)$.
}
\end{figure}

\begin{figure}
\centerline{\vbox{
\psfig{figure=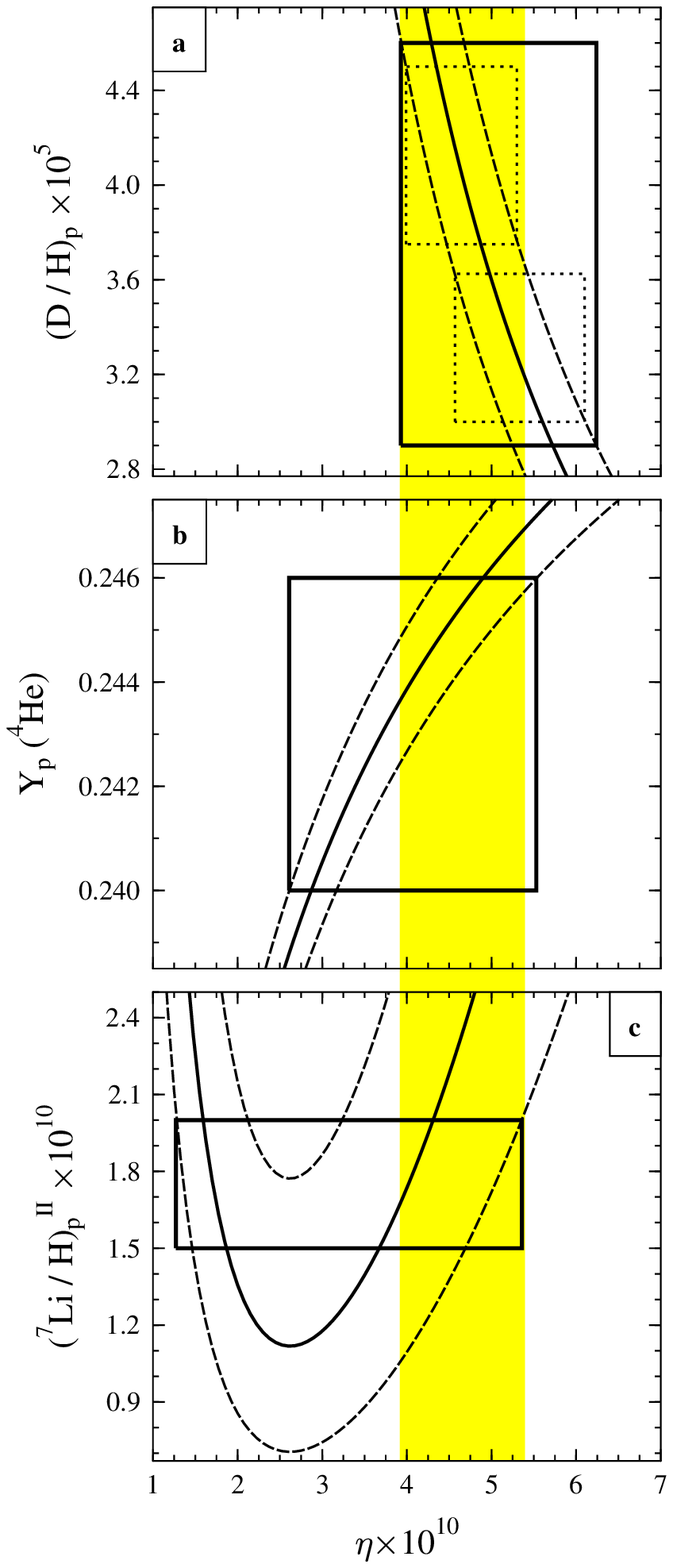,height=12.0cm,width=12.0cm}
}}
\vspace{-1.0cm}
\caption[]{
Comparison of predicted primordial abundances 
(denoted by the subscript p) with observational bounds. 
The theoretical BBN \ D, $^4$He, and $^7$Li yields (solid curves) and their uncertainties
(dashed curves) as function of $\eta$ (the baryon to photon number ratio) are computed by
using the parameterization given in \cite{S}. 
The abundances of D and $^7$Li are number ratios,
whereas Y$_p$ is the mass fraction of $^4$He. 
The hight of the solid-line rectangles in panels
{\bf b} and {\bf c} give the bounds from recent measurements of extra-galactic Y$_p$ 
\cite{ITL} and $^7$Li in population II (halo) stars 
\cite{BM}. The horizontal widths give the range of $\eta$ compatible
with the measurements. In panel {\bf a} the
solid-line rectangle corresponds to the 
uncertainty region for D/H towards Q1009+2956,
as calculated by the RMC procedure for an arbitrary 
velocity field configuration \cite{L6}.
The upper and lower
dotted-line rectangles are defined by the confidence regions  
(computed in \cite{L6}) for the fixed velocity field structures 
shown in Fig.~1({\bf c2}) and Fig.~1({\bf a2}), respectively.
The shaded region is the window for $\eta$ common to all results. 
}
\end{figure}
\vfill
}

\begin{iapbib}{99}{
\bibitem{L1} Levshakov S. A., Kegel W. H., 1994, \mn, 271, 161 
\bibitem{La} Levshakov S. A., 1995, Space Sci. Rev., 74, 285
\bibitem{L2} Levshakov S. A., Kegel W. H., 1996, \mn, 278, 497 
\bibitem{L3} Levshakov S. A., Takahara F., 1996, \mn, 279, 651 
\bibitem{Lb} Levshakov S. A., Takahara F., 1996, Astron. Lett., 22, 491
\bibitem{L4} Levshakov S. A., Kegel W. H., 1997, \mn, 288, 787 
\bibitem{L5} Levshakov S. A., Kegel W. H., Mazets I. E., 1997, \mn, 288, 802 
\bibitem{L6} Levshakov S. A., Kegel W. H., Takahara F., 1997, \mn, (submit.)
\bibitem{VO} van Ojil R. \et, 1997, \aeta, 317, 358
\bibitem{BT} Burles S., Tytler D., 1996, preprint astro-ph/9603070
\bibitem{ITL} Izotov Yu., Thuan T. X., Lipovetsky V. A., 1997, ApJS, 108, 1
\bibitem{BM} Bonifacio P., Molaro P., 1997, \mn, 285, 847
\bibitem{S} Sarkar S., 1996, Rep. Prog. Phys., 59, 1493
}
\end{iapbib}
\end{document}